\documentclass{IEEEtran}
\usepackage{cite}
\usepackage{amsmath,amssymb,amsfonts}
\usepackage{graphicx}
\usepackage{tabularx}
\usepackage{booktabs}
\usepackage{algorithm,algpseudocode}

\newcolumntype{Y}{>{\centering\arraybackslash}X}

\newcommand{\xtone}[1]{\mathbf{x}^{#1}_{t_{k+1}}}

\begin{document}
\title{Generative-Model-Based Fully 3D PET Image Reconstruction by Conditional Diffusion Sampling}
\author{George Webber, Yuya Mizuno, Oliver D. Howes, Alexander Hammers, Andrew P. King and Andrew J. Reader
\thanks{\scriptsize Presented at IEEE NSS MIC RTSD 2024, see doi: 10.1109/NSS/MIC/RTSD57108.2024.10657861. Copyright (c) 2024 IEEE.}
\thanks{\scriptsize George Webber would like to acknowledge funding from the EPSRC Centre for Doctoral Training in Smart Medical Imaging [EP/S022104/1] and via a GSK Studentship.}
\thanks{\scriptsize George Webber (e-mail: george.webber@kcl.ac.uk), Andrew P. King and Andrew J. Reader are with the School of Biomedical Engineering and Imaging Sciences, King’s College London, UK. Yuya Mizuno and Oliver D. Howes are with the Institute of Psychiatry, Psychology and Neuroscience, King’s College London, UK. Alexander Hammers is with the King’s College London \& Guy's and St Thomas' PET Centre.}
\vspace{-0.2cm}
}

\maketitle

\begin{abstract}
Score-based generative models (SGMs) have recently shown promising results for image reconstruction on simulated positron emission tomography (PET) datasets.
In this work we have developed and implemented practical methodology for 3D image reconstruction with SGMs, and perform (to our knowledge) the first SGM-based reconstruction of real fully 3D PET data.
We train an SGM on full-count reference brain images, and extend methodology to allow SGM-based reconstructions at very low counts (1\% of original, to simulate low-dose or short-duration scanning).
We then perform reconstructions for multiple independent realisations of 1\% count data, allowing us to analyse the bias and variance characteristics of the method.
We sample from the learned posterior distribution of the generative algorithm to calculate uncertainty images for our reconstructions.
%We evaluate the method's performance on real full- and low-count PET data and compare with conventional OSEM and MAP-EM baselines, showing that our results outperform those of the baseline methods both visually and in terms of bias-variance performance.
We evaluate the method’s performance on real full- and low-count PET data and compare with conventional OSEM and MAP-EM baselines, showing that our SGM-based low-count reconstructions match full-dose reconstructions more closely and in a bias-variance trade-off comparison, our SGM-reconstructed images have lower variance than existing baselines.
%We also sample from the learned posterior distribution of the generative algorithm to calculate uncertainty images for our reconstructions.
%The generative-model based algorithm can generate multiple reconstructions to sample from the learned posterior distribution and hence calculate uncertainty images for our reconstruction.
Future work will compare to supervised deep-learned methods, with other avenues for investigation including how data conditioning affects the SGM's posterior distribution and the algorithm's performance with different tracers.

\end{abstract}

\begin{IEEEkeywords}
Positron Emission Tomography, Image Reconstruction Algorithms, Deep Learning, Generative AI
\end{IEEEkeywords}

\vspace{-0.2cm}
\section{Introduction}
\label{sec:introduction}
%Low-count PET acquisition is desirable for patient safety and maximising scanner throughput. 
Low-count positron emission tomography (PET) data arises in contexts such as low-dose administration and acquisition time reduction.
However, the inverse problem of PET reconstruction from low-count data suffers from high-variance Poisson noise.
Recently proposed deep learning methods incorporate learned prior information into the reconstruction process to mitigate this issue.
%This also allows SGM-based reconstruction to remain agnostic to the specifics of the scanner setup.  
Score-based generative models (SGMs) are state-of-the-art generative models that only require \textit{unpaired} high-quality images for unsupervised training, thus decoupling scanner-specific considerations from the training process for greater generalisability \cite{chung_score-based_2022}.
State-of-the-art reconstruction has been shown for MR and CT with SGMs, while Xie et al. and Singh et al. have shown state-of-the-art simulated results for PET \cite{xie_joint_2023} \cite{singh_score-based_2024}.
In this work, we develop and investigate state-of-the-art SGM-based fully 3D image reconstruction methodologies on very low count real PET data, using the example of \textit{in vivo} $[^{18}$F]DPA-714 distributions.

\section{Theory}
PET image reconstruction may be formulated as an inverse problem by modelling the mean $q$ of our measurements $m$ as \[ q = Ax + b\] where $x$ represents the radiotracer distribution, $ A $ represents our system model and $ b $ models scatter and randoms components. Let $ p(x) $ be the prior probability density of image $ x $.

The SGM framework involves two stochastic processes, a forward ``diffusion'' process that maps from a distribution of images to a standard high-dimensional Gaussian, and its reverse, the backward process which we seek to learn. Assuming access to the score function $ \nabla \log(p(x)) $, the backward process may be expressed analytically and numerically solved, thereby allowing us to sample from the image distribution.

We train an unconditional time-dependent SGM $ s_{\theta}(x, t) $ to model the score function from a training set of high-quality 2D transverse brain slices extracted from 3D full-count images, via denoising score matching \cite{song_generative_2019}.

\begin{figure*}[t]
    \vspace{-0.7cm}
	\centering
    \begin{center}
    	\includegraphics[width=\textwidth]{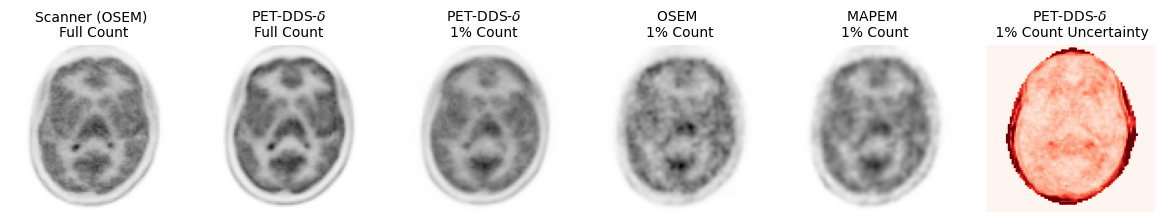}
    \end{center}
    \vspace{-0.5cm}
	\caption{\footnotesize Example transverse slices from 3D images reconstructed from a test dataset. The uncertainty image (far-right) is a preliminary result from sampling the posterior by repeating 25 randomly-seeded PET-DDS-$\delta$ reconstructions, and calculating the voxel-wise coefficient of variation.}
	\label{fig:example_images}
\end{figure*}

To condition on fully 3D PET sinogram data, we alternate steps solving the backwards-time generative process with steps to encourage data consistency. One method that does this is Singh et al.'s PET-DDS, an adaptation of Decomposed Diffusion Sampling to the case of non-negative PET images with high dynamic range \cite{singh_score-based_2024}. In PET-DDS, data consistency steps constitute gradient descent on the standard Poisson log-likelihood $L_j$ for subset $j$, an axial relative difference prior (RDP) to enable 3D reconstruction and a third term to prevent straying too far from the diffusion output. As we consider reconstruction from ${\sim}10\times$ fewer counts than Singh et al., our log-likelihood gradients are significantly larger. Therefore, we extend this algorithm to give PET-DDS-$\delta$, by introducing an additional hyperparameter $ \delta $ to moderate the rate of gradient descent ($GD$) towards the reconstruction objective.
%Also, to further prevent over-convergence to the noisy maximum likelihood estimate, the final 20\% of iterations are not conditioned on measurements.
Let our forward process be defined by $ p(\mathbf{x}_t|\mathbf{x}_0) = \mathcal{N}(\mathbf{x}_t; \gamma_t\mathbf{x}_t, \nu_t^2 \mathrm{I})$. Let stochasticity be $\eta_t$. Then for each $k$ of $N=100$ iterations, given iterate $\xtone{}$:
\begin{align*}
    &\mathbf{x}^{(s)}_{t_k} \leftarrow s_{\theta}(\xtone{}, t_{k+1})  \hspace{1.74cm} \rhd \text{ Diffusion step} \\
    &\xtone{0} \leftarrow \gamma_{t_{k+1}}^{-1} \cdot (\xtone{} + \nu_{k+1}^2 \mathbf{x}^{(s)}_{t_k}) \hspace{0.5cm} \rhd \text{ Tweedie estimate (de-noise)}\\
    &\Phi_j(\xtone{i}) \leftarrow L_j(| \xtone{i}) + \frac{1}{n_{\text{sub}}} \bigl( \lambda^{\text{RDP}} RDP_z(\xtone{i}) \\
    & \hspace{1.6cm} - \lambda^{\text{DDS}} ||\xtone{i}  - \xtone{0}||^2_2 \bigr)  \hspace{0.7cm} \rhd/\triangledown \text{  Data consistency}\\
    &\textbf{for } i = 1,...,p\textbf{ do } \xtone{i} \leftarrow GD(\delta \cdot \Phi_{j+i}(\xtone{i-1}))\\
    & j \leftarrow j+i ; \hspace{0.5cm} \mathbf{z} \sim \mathcal{N}(0, \mathrm{I}) \hspace{1.65cm} \hspace{0.7cm} \triangledown \text{ Re-noising step}\\
    &\mathbf{x}_{t_k} \leftarrow \gamma_{t_k} \xtone{p} - \nu_{t_{k+1}}\sqrt{\nu_{t_k}^2 - \eta_{t_{k+1}}^2}\mathbf{x}^{(s)}_{t_k} + \eta_{t_{k+1}} \mathbf{z}
\end{align*}

\section{Experiments}
Fifty-five static $[^{18}$F]DPA-714 brain datasets (from the Inflammatory Reaction in Schizophrenia team at King's College London) were used for SGM training and validation \cite{muratib_dissection_2021}. Data were acquired from 1-hour scans with the Siemens Biograph mMR, with approximately 200 MBq administered, with total counts in the range $2.9 \times 10^8 - 1.3 \times 10^9$. At full-count, high-quality images (voxel size 2 mm $\times$ 2 mm $\times$ 2 mm ; 3D image size $128 \times 128 \times 120$) were reconstructed with the scanner defaults (OSEM with 21 subsets and 2 iterations). An SGM was trained for 100 epochs (a value identified via 5-fold cross-validation) to learn the score function of these data.

An additional 7 datasets were reserved for validation and testing of the reconstruction process.
For our SGM reconstruction method we used PET-DDS-$\delta$, with 100 diffusion steps and 5 reconstruction objective steps per diffusion step (i.e. $p=5$).
For comparison, we implemented OSEM and MAP-EM (with patch-based regularisation \cite{wang_penalized_2012}).
% and the method FBSEMnet for deep learning unrolled iterative reconstruction \cite{mehranian_model-based_2020}
Each method used the same ParallelProj 3D projector \cite{schramm_parallelprojopen-source_2024} with span 11 axial compression applied.
To obtain low-count sinograms, prompts and randoms were sampled at 1\% of counts assuming independent Poisson statistics, with smoothed randoms and scatter sinograms re-estimated using scanner software.

For PET-DDS-$\delta$, reconstruction hyperparameter values of $ \lambda^{\text{RDP}} = 10^{-4} $, $\delta=0.2$, $\lambda^{\text{DDS}} = 1.0$ were selected to minimise NRMSE on the validation dataset.
On this dataset, reconstructions were performed on 20 independent realisations of 1\% count data, allowing calculation of bias and variance.
The remaining 6 datasets were used to evaluate methods at their optimal hyperparameters, with the full-count scanner reconstructions taken as ground truth.

\section{Preliminary Results and Discussion}
\begin{figure}[t]
	\centering
	\includegraphics[width=\linewidth]{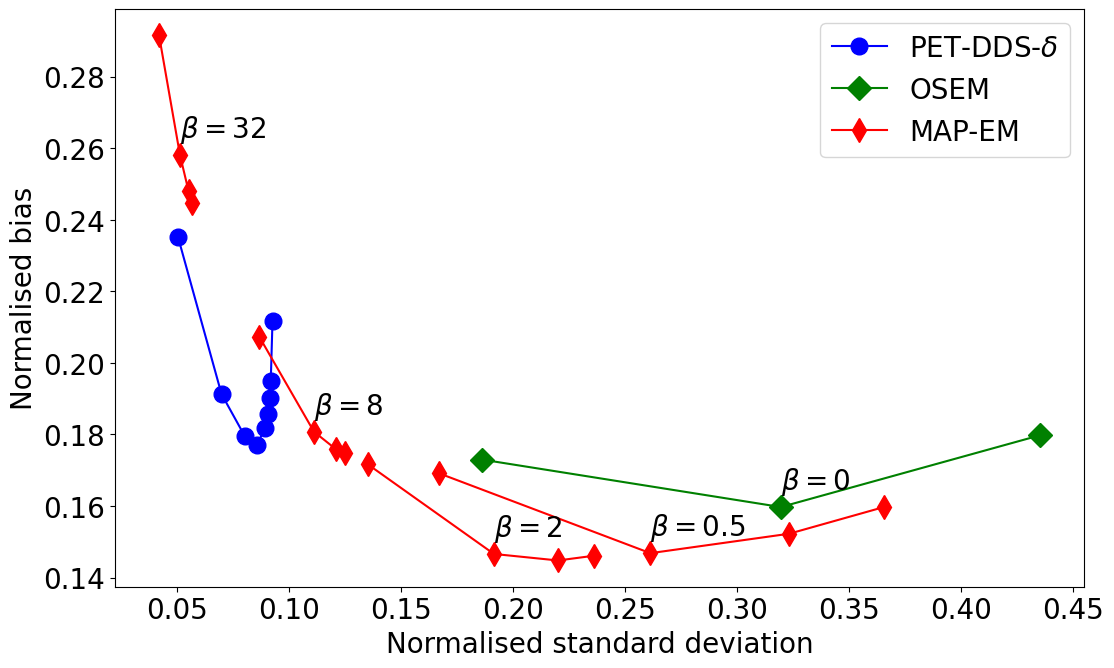}
    \vspace{-0.7cm}
	\caption{\footnotesize Bias-variance trade-off of reconstruction algorithms. For MAP-EM, iteration numbers (from 21 to 84) and regularisation strengths $\beta$ (from 0 to 32) were used. For PET-DDS-$\delta$, $ \delta $ varied from 0.05 to 0.8.}
	\label{fig:bias_variance}
\end{figure}

Table \ref{tbl:main_results} shows that PET-DDS-$\delta$ of the 1 \% data more closely matches the full-count reconstruction, achieving lower NRMSE values compared to the baseline methods as expected. This analysis is supported qualitatively by Figure \ref{fig:example_images}, in which PET-DDS-$\delta$ retains higher visual fidelity compared to the other low-count reconstructions. In our bias-variance analysis in Figure \ref{fig:bias_variance}, we see that the reduced NRMSE values are primarily due to reduced variance with respect to the Poisson noise realisations, with a higher minimum bias observed for PET-DDS-$\delta$ than for MAP-EM.

\begin{table}[t]
	\caption[]{\footnotesize Average reconstruction performance from 1\% count data over 6 test datasets. Ground truth was taken as the full-count scanner reconstructions}
    \vspace{-0.4cm}
	\label{tbl:main_results}
	\begin{center}
		\begin{sc}
			\begin{tabularx}{\linewidth}{l|YYY}
	\toprule
	Metric & NRMSE (\%) $\downarrow$ & PSNR (dB) $\uparrow$ & SSIM (\%) $\uparrow$ \\
	\midrule
	OSEM     & $26.1 \pm 3.1$ & $28.4 \pm 3.3$ & $74.2 \pm 7.6$ \\
	MAP-EM   & $23.7 \pm 1.3$ & $29.2 \pm 2.4$ & $77.6 \pm 6.1$ \\
	%FBSEMnet &  $22.8 \pm 1.6$ & $29.6 \pm 2.9$ & $77.2 \pm 6.7$ \\
    PET-DDS-$\eta$  & $\textbf{21.2} \pm 1.6$ & $\textbf{30.1} \pm 2.9$ & $\textbf{79.4} \pm 5.7$ \\
	\bottomrule
\end{tabularx}
		\end{sc}
		\vskip -0.4in
	\end{center}
\end{table}

\section{Summary}

We have developed and extended practical methodology for 3D image reconstruction from low-count PET data with SGMs, and performed SGM-based reconstruction of real fully 3D PET data. Future work will compare to other deep-learned methods, and may also focus on: investigating the relationship between the data conditioning and the SGM's image manifold; conditioning on MR information; and, investigating the algorithm’s performance with $[^{18}$F]FDG data.

\appendices

\bibliography{IEEE_MIC_Abstract_1_refs}

\end{document}